# Comparison of the P300 detection accuracy related to the BCI speller and image recognition scenarios

S.A. Karimi [1], A.M. Mijani[2], M.T. Talebian[1] and S. Mirzakuchaki[*1]

1. Electronic Research Institutes (ERI), Department of Electrical engineering, Iran University of science & technology, Tehran, Iran

2. BiSIPL, Department of Electrical Engineering, Sharif university of Technology, Tehran, Iran.

**Abstract**
There are several protocols in the Electroencephalography (EEG) recording scenarios which produce various types of event-related potentials (ERP). P300 pattern is a well-known ERP which produced by auditory and visual oddball paradigm and BCI speller system. In this study, P300 and non-P300 separability are investigated in two scenarios including image recognition paradigm and BCI speller. Image recognition scenario is an experiment that examines the participants' knowledge about an image that shown to them before by analyzing the EEG signal recorded during the observing of that image as visual stimulation. To do this, three types of famous classifiers (SVM, Bayes LDA, and sparse logistic regression) were used to classify EEG recordings in six classes problem. Filtered and down-sampled (temporal samples) of EEG recording were considered as features in classification P300 pattern. Also, different sets of EEG recording including 4, 8 and 16 channels and different trial numbers were used to considering various situations in comparison. The accuracy was increased by increasing the number of trials and channels. The results prove that better accuracy is observed in the case of the image recognition scenario for the different sets of channels and by using the different number of trials. So it can be concluded that P300 pattern which produced in image recognition paradigm is more separable than BCI (matrix speller).

**Keywords:** *EEG, P300, BCI, Classification, Channel number effect.*

**1. Introduction**

The Electroencephalography (EEG) signals are generated by the firing of neurons in the brain and their related electrical activities. German psychiatrist Hans Berger, coined the name Electroencephalogram after he could successfully record this specific activity of the brain in 1924 [1]. This electrophysiological monitoring method is a non-invasive procedure of the brain activity recording, which has several advantages over similar methods such as its low cost and ease of the operation [2].

After the specific visual and auditory signals stimulate the brain, certain patterns which are called Event-related Potential (ERP) are recorded by the EEG. One well-known ERP is P300 or P3 which generated as a result of the induction with a rare and meaningful stimulus, called "oddball" [3]. It is interpreted as a fast response of the central nervous system to the stimulus that is meaningful or in other words; the participant is familiar with it. It is an unconscious response; thus, it might hardly be altered with the participants' intention [4].

It is a positive-going wave that occurs approximately 300 milliseconds after the stimulation with a clear scalp distribution, in which the most significant value is typically at the Pz electrode and the smallest one at Fz. The intermediate value lies at Cz channel in a standard 10-20 system of the recording EEG [5].

Image recognition scenario is an experiment that examines the participants' knowledge about an image that shown to them before by analyzing the EEG signal recorded during the observing of that image as a visual stimulation. Several kinds of research have been done in order to detect the target image among an image string using p300 component, that this paradigm was called rapid serial visual presentation (RSVP) [6-8]. RSVP paradigm in addition to image search applications applied in the design of p300 speller system [9, 10].

Brain-computer interface that is abbreviated as BCI in the literature is a system that creates a relationship between the brain and a computer. It means that a BCI system could discover the processes done in the brain and do the appropriate tasks according to that by the computer. Such a system is predominantly utilized in helping the disabled, especially people with motor disabilities [9]. EEG is commonly used in the BCI systems because of its non-invasive nature and its direct relationship with the brain function. By analysis of the EEG recordings, a computer can recognize the brain's intention and making commands for the output devices. P300 is a useful pattern trough EEG signal to this end because it is unconscious and it directly reflects the brain intentions [11].

One of the most common BCI systems that are called BCI speller is a mechanism that helps the subject to type text by considering the letters of the words. Recently various studies around the world have focused on investigating P300 applications and development of the BCI systems. BCIs have a variety of applications, one being Speller systems. The first Speller paradigm based on ERP was the matrix Speller which was introduced by Farwell and Donchin [12]. Research has proven that Matrix Speller has a disadvantage and it is gaze dependent [9]. Many researchers attempted to overcome this issue, and their results led to changing the type of modality, and using audio [13] and touch [14] instead of vision.

Erwei Yin et al. [15] proposed a novel hybrid BCI approach for increasing the spelling speed. In this approach, the P300 and the visually evoked steady-state potential (SSVEP) detection mechanisms are devised and integrated in order to make use of two brain signals possible for spelling at the same time. The results obtained for the fourteen healthy subjects demonstrate that the average online practical information transfer rate, including the time of the break between selections and error corrections, achieved using this approach was 53.06 bits/min [15].

Riccio et al. [16] studied the relative improvement of two modalities of BCI system control; a P300-based and a hybrid P300 electromyographic-based mode of control. They consider eleven persons, eight healthy ones and the other three with the motor disability. Evaluation of the system's function was revealed in three categories, namely, system accuracy, throughput time and satisfaction of the user.

In both groups, higher marks were in the hybrid P300 electromyographic-based mode. They used data driven by electromyographic recording as a correction tool [16].

Akram et al. [17] in a study in 2015 proposed a Text on nine keys (T9) based on the word typing systems relying on the P300 BCI system. For the sake of usability, they integrated a smart dictionary into the system for word suggestion and used a Random Forest classifier for the enhancement of the accuracy. Their results showed 51.87 percent improvement in the typing speed compared with the conventional systems and higher accuracy for their classifier [17].

Halder et al. tried to make a solution based on the P300 BCI system for control of the web-browser and multimedia player. They used dynamic matrices for accessing higher speed and evaluated their system by both healthy and end-user persons. They presented a web browser, and a multimedia user interface adapted for control with a brain-computer interface (BCI) which may benefit severely motor disabled individuals. The web browser dynamically determines the most efficient P300 BCI matrix size to select the links on the current website and the multimedia player corresponding to the piece of software in hand. The results reveal that healthy participants experience 90% of accuracy in media player and 85% in web browsing predefined tasks, while disabled participants report 62% and 58% accuracy for abovementioned tasks [18].

Haghighatpanah et al. worked on a single trial, single-channel algorithm. They used wavelet decomposition and ICA method for the feature extraction and LDA algorithm as the classifier. Reportedly, they could achieve 65% detection accuracy [19].

Another application of the P300 signal that has been continuously discussed in the literature is lie detection or in a better expression, the deception detection. As discussed before, the rising of P300 pattern in the EEG recordings shows a sign of familiarity or intention of the participant with the stimulation. This concept is the base of all applications of the P300 signal. In lie detection applications, we can review an image detection scenario because the stimulation is often visual and the goal is to find the knowledge of participant about the image shown as the stimulation. One of the most critical problems in lie detection applications is designing the best scenario for the test. For achieving this goal, this study was done to compare two primary types of stimulation that use in P300 analyzing based studies. By comprising these methods, we can conclude that which factor involving in the creation of P300 is more important and a better designing of test scenario for deception detection could be obtainable.

Wang et al. in a study in 2013 created a lie detection system based on the P300 by developing a simple and feasible hierarchical knowledge-base construction and test method. In this study, each subject was asked to provide to the experimenter five numbers (all four digits long), one of them being their year of birth. In each test, each number was displayed to the subject randomly with thirty repetitions as stimulations. This type of stimulation is like BCI application. They describe how a hierarchical feature space was formed and which level of the feature space was sufficient to accurately predict concealed information from the raw EEG signal in a short time. The results indicate a high accuracy of 95.23% in recognizing concealed information with a single EEG electrode within about 20 seconds [20].

Yijun et al. in research in 2014 introduced a classification method based on the ICA and ELM for using in lie detection based on P300. The participants were randomly divided into two groups: the liar group and truth-teller group. The images of six watches with different characteristics were prepared. A box containing two watches was given to the guilty. Then, the guilty were instructed to steal one, which was served as the probe (P) stimuli. The other objects in the box are the Target (T) stimulus. The remaining four objects are irrelevant (I) stimuli. For the innocent, they only saw one watch (T stimulus) in the box and stole nothing. The standard three stimuli protocol was employed in this study. First, they used ICA for identifying the P300 ICs. The features were extracted from time and frequency domain and finally, they used these features to train three kinds of classifiers: Extreme learning machine (ELM), Backpropagation neural network (BPNN) and Support vector machine (SVM). Their reports suggest that the method combining ICA with ELM has the best result in terms of the accuracy [21].

## 2. Methods

One of the controversial subjects in the field of P300 detection is the relationship between the type of stimulation and the clarity of the P300 pattern in the EEG. Does it mean that which type of stimulation can better and stronger create P300? In this study, we try to find an answer to this question. Two factors are involved in the creation of P300: first, the familiarity of stimulation with the participant, and second, having consideration from the participant about the concept of the stimulation. For the first case consider a person that rubes a special wallet, when he looks at the picture of that special wallet, like the stimulus, in a manner, P300 wave in his brain appears. For the second case, consider a disabled person willing to type a special letter by a P300 based BCI system, when the picture of that letter is shown for him as the stimulation in a particular algorithm, P300 will appear in the EEG recording. This shape is not particularly special, but the concept of the letter and intention of the disabled person make it special for him.

### 2.1. Data Sets

The first dataset is the EPFL image recognition database [22]. In this study, five disabled and four healthy individuals were examined. Disabled persons were limited with the wheelchair but have the varying ability to control their muscles and communications. Subjects 1 and 2 were able to perform simple, slow movements with their arms and hands but were unable to control other extremities. Spoken communication with subjects 1 and 2 was possible, although both subjects suffered from mild dysarthria. Subject 3 was able to perform restricted movements with his left hand but was unable to move his arms or other extremities. Spoken communication with subject 3 was impossible. However, the patient was able to answer yes or no questions with eye blinks. Subject 4 had very little control over arm and hand movements. Spoken communication was possible with subject 4, although mild dysarthria existed. Subject 5 was only able to perform extremely slow and relatively uncontrolled movements with hands and arms. Due to severe hypophonia and large fluctuations in the level of alertness, communication with subject 5 was challenging. Sixth to ninth participants were Ph.D. students working in the lab (all men were in the ages (30 ± 2/3)). None of these persons had a known nerve dysfunction. Each person was tested four times, and an image was shown every time. The brain waves of these individuals were recorded based on P300. The first two sessions took place one day and two subsequent sessions on the other day. For all participants, the time between the first and last session was less than two weeks. Each of these sessions consisted of six runs each included multiple random presentations of six images. Before each session, participants were asked to select a particular picture and count the number of flashing that through the random presentation. During the stimulations, EEG recording was done from 32 channels according to the standard 10-20 system. Counting was intended to monitor the degree of attention from the participants to the examination [22].

The second database analyzed in this study is the BCI2003 mental typing competition database (Data set IIb). Institute of Berlin Brain-Computer Interface competes with other institutions to develop in the context of BCIs. During each competition, several datasets are provided to the participants that they must answer certain questions with analyzing them. The data of these competitions is available for free on the website (http://bbci.de/competition). One of these datasets is related to the competition that has been held in 2003 named, Berlin BCI Competition II-Data set IIb. This dataset includes a user's EEG signal that was recorded while using a P300 speller [23].

The BBCI Dataset IIb data consists of three EEG recording sessions, during which a subject (which was the same in all three sessions) has typed some words by the P300 Speller. The P300 Speller is the same as one that was first introduced in 1988 by Donchin and Fawell, then described in 2000 by Donchin [24]. In this P300 speller, the subjects viewed a display of a 6*6 matrix filled with alphabets and numbers. The characters were presented as white characters on a black background, using a moderate and easily visible intensity. The subjects were instructed to observe the display and to count the number of times the row, or the column, containing the designated target letter was intensified. The rows and the columns were intensified in a random sequence in such a manner that all six rows and six columns were intensified before any was repeated. A "trial" in the study is thus defined as the intensification of all 12 elements of the matrix. Simultaneously the EEG was recorded from tin electrodes in an electrode cap according to the standard 10-20 system and right mastoid sites, referred to the left mastoid.

The analysis in this study was done by using the data of 4, 8, and 16 channels. In 4 channel analyzing, data from Fz, Cz, Pz, and Oz was chosen and in 8 channel analyzing data from P3, P4, P7, and P8 were added and in 16 channel analyzing data from FC1, FC2, CP1, CP2, C3, C4, O1, and O2 were added according to standard 10-20 system. See the figure 1.

**2.2. Features extraction**

Preprocessing and feature selection for both databases are done through the following eight steps:

1. Reference determination: The average signal from two mastoid electrodes was used as the reference.
2. Filtering: The third-order bandpass butter-worth filter was used for filtering the data. The cut-off frequency was set to 1 Hz and 12 Hz. The MATLAB butter-worth function was used to calculate the filter coefficients.
3. Downsampling: After passing the signal from the bandpass filter, the sampling rate should be reduced to 32 Hz, so that the calculations and machine learning algorithms could be operational. In this study, because the initial sampling rate is 2048, then the sampling rate could be reduced by 64.
4. Epoch extraction: Length of each epoch considered 1000 (ms) started definitely after the stimulation onset.
5. Winsorizing: For rejecting the artifact effects, the values of ten percent up and down the range of domain were saturated by the value of ninety and ten percent respectively.
6. Normalizing: The samples from each electrode were scaled to lie in the range of [-1, 1].
7. Electrode selection: Four electrode structures with different numbers of electrodes were tested.
8. Building feature vector: Samples of selected electrodes were added to the feature vector. The dimensions of the feature vectors were $N_t*N_e$ where $N_e$ is the number of electrodes that can be 4, 8, or 16 according to type of channel selection and $N_t$ represents the number of time samples in a test that according to present datasets was 32.

Due to the duration of epochs in the databases under discussion, which is 1000 (ms) and the resulting down sampling rate of 32 Hz, $N_t$ is always equal to 32. Depending on the configuration of the electrodes, $N_e$ is 4, 8 or 16.

**2.3. Selected classifier**

In this study for classifying the separated and preprocessed epochs to P300 included or without P300 we use three kinds of classifiers as can be seen in the following. The main challenges in this application are the low level of SNR and the variability of the ERP pattern between individuals and at different times. The classification step is essential and selecting an appropriate classifier according to the type of system, and of course, the existing data can have a substantial effect on the overall system efficiency.

The first classifier that has been used in this study is Bayesian linear discrimination analysis (Bayes LDA). This classifier is the approach of using the Bayesian Formulation in the LDA, which is opposed to the Fischer-based approach. In short, in this classifier, in a K class problem, we seek to establish a linear separator for each class, so that for each sample belonging to the class, the term of that sample will be maximized. Suppose that there are (x, y) in which x represents input or sample vectors and y represents the labels of classes. According to the Bayesian formulation we have:

$$p(y_k|x) = \frac{p(x|y_k)p(y_k)}{p(x)} \propto p(x|y_k)p(y_k)$$

Given Gaussian distribution for the probability of samples in each class, the relationship will be as follows

$$= ln\left[\frac{1}{(2\pi)^{p/2}|\Sigma|^{1/2}} \exp\left(-\frac{1}{2}(x-\mu_k)^T \Sigma^{-1}(x-\mu_k)\right)\right]$$

This specifies a linear function in separating classes from each other [22].

The second classification method used here is the support vector machine (SVM). SVM is a prediction algorithm used in the classification and regression problems. SVM has long been developed and is a combination of computational theories such as margin hyperplane and kernel. In other words, SVM is a technique used to obtain the most probable hyperplane to separate two classes. It is done by measuring the hyperplane's margin and determines its maximum point. The margin is defined as the distance between the corresponding hyperplane and the nearest pattern from each class. Moreover, this nearest pattern is called support vector. Meanwhile, SVM can also be used to separate non-linear data.

The third classifier that has been used in this study is the generalized linear method (GLM). The problem of the generalized linear model is defined for the Lasso problem or the elastic net. The definition of these two problems is as follows:

For non-negative value of λ, Lasoo solves the following problem:

$$\min_{\beta_0,\beta}\left(\frac{1}{N} Deviance(\beta_0,\beta) + \lambda \sum_{j=1}^{p}|\beta_j|\right)$$

$\beta_0, \beta$ coefficients will fit the deviance of model. Deviance depends on the response of the model by fit coefficients. Deviance formulation is also dependent on the distribution is given to the lassoglm function. Minimizing the deviance that is penalized by the phrase λ is the same as maximizing the log-likelihood term that is penalized by the phrase λ. Deviance depends on the response of the model by fit coefficients. The Deviance formulation is also dependent on the distribution is given to the lassoglm function. Minimizing the deviance that is penalized by the phrase λ is the same as maximizing the log-likelihood term that is penalized by the phrase λ. N is the number of observations and λ acts as the regulatory parameter of the function. The parameters of the scalper vectors are the length of p. $\beta_0, \beta$ are the scalar vectors with the length of p. N is the number of observations and λ is the regulatory parameter of the function.

Due to the variable and noisy nature of P300 pattern, it is recommended to find classification methods that have the appropriate performance with a small number of training data.

### 2.4. Evaluation

To evaluate the performance of the classifier, we use the accuracy criterion. In this study, accuracy was calculated according to Hoffmann [22] method in EPFL image recognition database [22]. 10-fold cross validation used to train each classifier. Classifiers are trained to discriminate P300 and non-P300 patterns in two classes mode. Then each sequence of six images features is given to the classifiers and the images which have a maximum average score selected as target class. In the same way, every six rows and every six columns in BCI2003 competition were considered

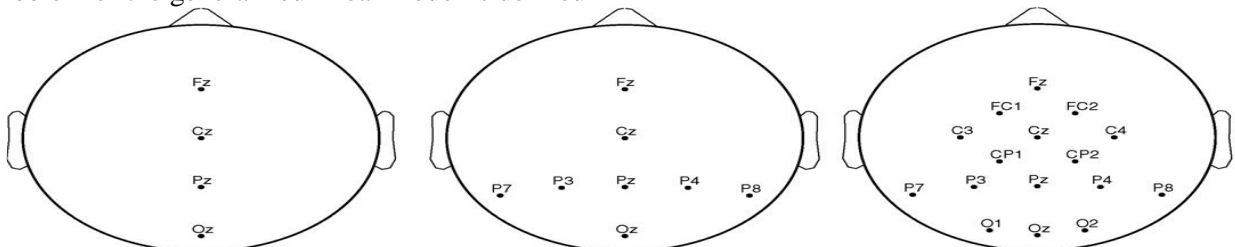

**Figure1- Configuration I (4 electrodes), configuration II (8 electrodes), and configuration III (16 electrodes).**

as a sequence of images in the EPFL data set. These rows and columns were trained and classified as EPFL data set. To compare two scenarios accuracy was used and the scenario which has better accuracy was considered a better scenario in producing P300.

## 3. Results

Figure 1 shows the classification accuracy with three methods of classification for data collected from EPFL image recognition database by eight electrodes. This result was averaged over all participants. The horizontal axis in the figure shows the number of stimulation trials. As can be seen in the picture increasing the number of probe stimulation clearly increases the classification accuracy. All three methods used in this study reach to an acceptable accuracy level and between these methods; Bayesian LDA mildly has a better performance.

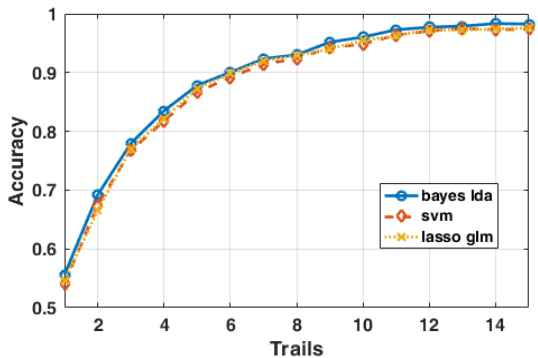

**Figure2- Classification accuracy with three methods for EPFL image recognition database by data from 8 electrodes.**

Figure 2 shows the relationship between the numbers of electrodes used and accuracy gained by the Bayesian LDA classifier. As the figure shows, by increasing the number of electrodes we gained better performance in classification and the gradient of accuracy by the trials also increased.

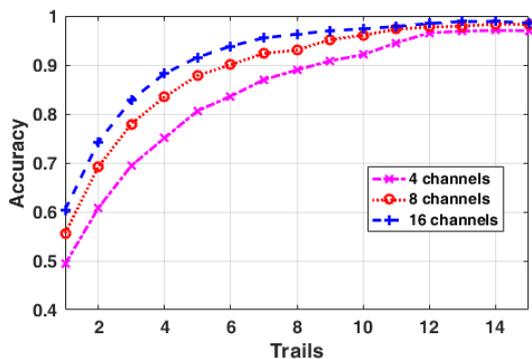

**Figure 3- Classification accuracy by the numbers of electrodes classified with Bayesian LDA for EPFL image recognition database.**

In the BCI2003 competition database, we have a six classes problem such as the former database. First, figure 3 can represent how is the relationship between the number of trials and accuracy for three different classification methods by using data gathered from the eight channels.

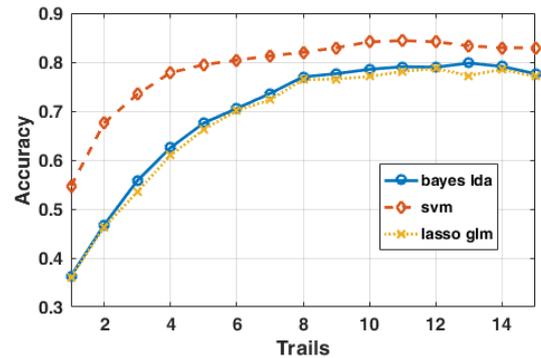

**Figure4- Classification accuracy with three methods for BCI2003 competition database by 8 channels.**

As it is evident in the picture accuracy will be increased by an increment in the number of trials but the maximum number of accuracy accessed is lower than it was for the former database. For this database, SVM works better than other classifiers. In Figure 4, you can see the relationship between increments in the number of channels by the accuracy gained with using of SVM method.

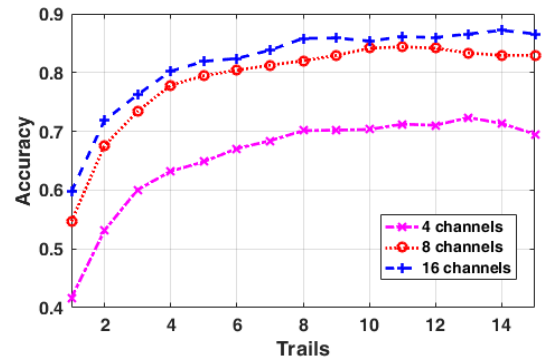

**Figure5- Classification accuracy by the numbers of channels classified with SVM for BCI2003 database.**

It is evident in the picture that increasing the channel number causes a better performance in classification.

Table 1 and 2 provide an appropriate condition for comparing the classification accuracy for two databases. Accuracies reported for EPFL image recognition database is based on the Bayes-LDA and for BCI2003 is based on the SVM because they have the best performance in the related database for classifying.

In Table 1, number of the trials considered to be five in each experiment and in Table 2, number of channels was eight.

**Table1- Comparison between the classification accuracy of databases with their best resulted classifier (trials=5).**

| database | EPF | BCI |
|---|---|---|

| channels | L | 2003 |
|---|---|---|
| 4 | 80 | 65 |
| 8 | 87 | 80 |
| 16 | 92 | 82 |
| max | 92 | 82 |

**Table2- Comparison between the classification accuracy of databases with their best resulted classifier (channels=8).**

| Database / Trials | EPFL | BCI 2003 |
|---|---|---|
| 2 | 75 | 67 |
| 5 | 93 | 79 |
| 10 | 99 | 85 |
| max | 99 | 82 |

## 4. Conclusion

It was shown that the increase in trials and averaging cause increment of accuracy significantly. Increasing the number of channels also has a positive effect on classification accuracy in both discussed databases. It was shown that the increase in trials and averaging cause increment of accuracy significantly. Increasing the number of channels also has a positive effect on classification accuracy. The best results were for P300 that was related to detecting image data. It has the most reliable type of P300 due to the need for recognizing the type of images. Subsequently, mental typing database had good results. We can conclude from these results that familiarity of stimulation with the participant, results stronger P300 than the P300 that is created by having consideration from the participant about the concept of the stimulation.